\documentclass[12pt] {article}
\usepackage{graphics}
\usepackage{epsfig}
\usepackage{psfrag}

\catcode`@=12 \topmargin -0.5in \evensidemargin 0.0in
\newcommand{\beq}{\begin{equation}}
\newcommand{\eeq}{\end{equation}}

\newcommand{\bea}{\begin{eqnarray}}
\newcommand{\eea}{\end{eqnarray}}

\newcommand{\ea}{{\it et al.}}

\def\lsim{\mathrel{\vcenter{\hbox{$<$}\nointerlineskip\hbox{$\sim$}}}}

\begin{document}
\thispagestyle{empty}
\begin{flushright}
UMD-PP-08-021\\
OSU-HEP-08-07\\
November, 2008\
\end{flushright}
\vspace{0.5in}
\begin{center}
{\LARGE \bf  Neutrino Mass Hierarchy and $n-\bar{n}$ Oscillations
from Baryogenesis\\}
\vspace{1.3in} {\bf
K.~S. Babu$^1$, P.~S. Bhupal Dev$^2$ and R.~N. Mohapatra$^2$\\}
\vspace{0.2in} {\sl $^1$ Department of Physics, Oklahoma State
University, Stillwater, Oklahoma 74078, USA\\} {\sl $^2$
Maryland Center for Fundamental Physics and Department of Physics, University of Maryland, College Park, Maryland
20742, USA\\} \vspace{1in}
\end{center}

\begin{abstract}
It has been recently proposed that the matter-antimatter asymmetry of the universe
may have its origin in ``post-sphaleron baryogenesis" (PSB). It is a TeV
scale mechanism that is testable at the LHC and other low energy experiments.
In this paper we present a theory of PSB within a quark-lepton
unified scheme based on the gauge group $SU(2)_L\times SU(2)_R\times
SU(4)_c$ that allows a direct connection between the baryon asymmetry
and neutrino mass matrix. The flavor changing neutral current constraints
on the model allow successful baryogenesis only for an inverted mass
hierarchy for neutrinos, which can be
tested in the proposed long base line neutrino experiments. The model
also predicts observable neutron--antineutron oscillation accessible to
the next generation of experiments
as well as TeV scale colored scalars within reach of LHC.

\vspace*{0.2in}

\end{abstract}

\newpage
\baselineskip 24pt



\section{Introduction}

Since Sakharov first suggested the three conditions that would have to be
satisfied by a microphysical theory to generate
matter--antimatter asymmetry of the universe~\cite{sakharov}, many beyond
the standard
model scenarios have been constructed for this purpose. The very earliest
ones that used proton decay in grand unified theories for this
purpose run into difficulty on several counts: first is that successful
inflation scenarios generally have reheating temperatures which are below
the generic baryogenesis temperatures, especially in the context of
supersymmetry, so that any GUT generated baryon number is erased by
inflation; secondly, if baryogenesis is
caused by $B-L$ conserving interactions as in $SU(5)$ models, they will be
destroyed by electroweak
sphalerons that are in equilibrium down to about 100 GeV.

In the mid 80's, a new mechanism was suggested that uses baryogenesis via
leptogenesis~\cite{lepto}. This mechanism is very attractive since it
arises within the framework of the seesaw mechanism~\cite{seesaw} that
explains small
neutrino masses. Here the initial lepton asymmetry is created far
below the GUT scale and is then converted by the electroweak
sphalerons~\cite{kuzmin} to a baryon asymmetry. This mechanism depends
crucially  on the properties of the electroweak sphaleron~\cite{kuzmin}
which serves as the source of $B$ violation. While this is one of the
most widely discussed schemes in literature today~\cite{sacha},
it may also have problems since adequate leptogenesis in
these models implies a lower bound on the leptogenesis scale~\cite{ibarra}
which is above the allowed reheating scale in supersymmetric
models~\cite{kohri}. It is also not so easy to test by low energy
experiments.

It is therefore
important to explore alternative mechanisms that can explain the
matter--antimatter asymmetry from particle decays around 100 GeV
temperatures which do not conflict with the above bounds on reheat
temperatures and at the same time
yield testable consequences at LHC and other low energy experiments.
Such a mechanism was proposed in two
recent papers~\cite{bmn1,bmn2}, where it was shown that with the use of higher
dimensional baryon violating operators, baryogenesis can occur
after the electro-weak sphalerons have gone out of thermal
equilibrium. This mechanism was called Post-Sphaleron Baryogenesis (PSB).

In this paper, we propose a theory for this mechanism within
a quark-lepton unified  gauge model for neutrino masses based on the
gauge  group $SU(2)_L\times SU(2)_R\times SU(4)_c$~\cite{ps} using the
symmetry breaking setup discussed in Ref.~\cite{marshak}. We show that in this
model, quark-lepton unification allows us to relate the baryon
asymmetry directly to neutrino masses via the type II seesaw
mechanism~\cite{type2}. The main result of our work is that
 successful baryogenesis can occur only for an inverted mass
hierarchy for neutrinos with a relatively large $\theta_{13}$.
Both of these predictions will be tested in ongoing  experiments
searching for neutrino-less double beta decay and neutrino oscillations.

The salient feature of PSB mechanism is that baryogenesis occurs
via the direct decay of a scalar boson $S_r$ having a weak scale
mass and a higher dimensional baryon violating coupling.  $S_r$ is
the real part of a baryon number carrying complex scalar $S$,
which acquires a vacuum expectation value (vev). In the context of the
$SU(2)_L\times SU(2)_R\times SU(4)_c$ model, $S_r$ is the real part
of a Higgs scalar field belonging to $(1,3,10)$ representation
of whose vev breaks the
$SU(2)_R\times SU(4)_c$ symmetry down to the $U(1)_Y$ of the standard
model. The decays
$S_r \rightarrow 6q^c$ and $S_r \rightarrow 6 \bar{q}^c$
 provide the source for $B$ asymmetry. When the $S$-field has a vev, the
decay process generates an interaction that causes neutron-antineutron
oscillation as shown in Ref.~\cite{marshak}. The parameter domain of our
theory where
adequate baryogenesis occurs predicts that neutron-antineutron
oscillation should occur at a rate observable in currently available reactor
facilities.

\section{Basic ingredients of Post-Sphaleron Baryogenesis and its
$SU(2)_L\times SU(2)_R\times SU(4)_c$ embedding}
A starting Lagrangian for PSB that gives rise to the higher dimensional
$B$-violating decay is given by~\cite{bmn1}
\begin{eqnarray}
{\cal L}_I &=& {h_{ij} \over 2}\Delta_{d^c d^c} d^c_id^c_j + {l_{ij}
\over 2} \Delta_{u^c u^c}u^c_iu^c_j + {g_{ij} \over 2} \Delta_{u^c
d^c} (u^c_id^c_j+u^c_jd^c_i) \nonumber \\
&&+{\lambda_1 \over 2} S \Delta_{u^c u^c}\Delta_{d^c
d^c}\Delta_{d^c d^c}+{\lambda_2 \over 2} S\Delta_{d^c
d^c}\Delta_{u^c d^c}^2 + {\rm h.c.}
\label{eq:1}
\end{eqnarray}
Here the $\Delta_{u^c u^c}$, etc. are color sextet scalar fields.
From the above equation, we see that when the scalar field $S$, which
has $B-L=2$ is given a vev, it leads to cubic scalar field
couplings of the type $ \Delta_{u^c u^c}\Delta_{d^c
d^c}\Delta_{d^c d^c}$ and $\Delta_{d^c
d^c}\Delta_{u^c d^c}^2 $ leading to baryon number violation by
two units.

We note that not all of the $( \Delta_{u^c u^c},  \Delta_{u^c
d^c},  \Delta_{d^c d^c} )$ fields are needed for $B$
violation and $n \leftrightarrow \bar{n}$ oscillation:  either $(
\Delta_{u^c d^c} , \Delta_{d^c d^c} )$
or $( \Delta_{u^c u^c},  \Delta_{d^c d^c} )$ pair will do. In fact,
consistency with flavor
 changing neutral current constraints and $n-\bar{n}$ oscillation
limits allow for
only two of these three scalar states to be light near the TeV scale.
The third state (in our case  $ \Delta_{u^c u^c} $, as we will see
below) will have mass of order 100 TeV.

Baryon asymmetry arises in this scheme from $W$-loop corrections to the $S_r$ decays and
is therefore directly linked to CKM mixing \cite{bmn1}.

The constraints on the parameter space of the model arise from the fact
that the decay of $S_r$ occurs below 100 GeV and above 200 MeV or so -- the
former to ensure that the sphalerons do not play any role in baryogenesis
and the latter so that quarks in the cosmic soup have not combined to form
hadrons, which will affect the decay estimates -- and from the fact that
the model must reproduce observed neutrino masses and mixings. If
 baryon asymmetry is created above the electroweak phase
transition temperature, all of the baryon asymmetry will be washed out
since there are both $B+L$ violating sphaleron interactions as well as $W_R$
mediated $\Delta L=2$ scatterings of right--handed Majorana neutrinos in
equilibrium at that temperature.

Before discussing the constraints on the parameters of the model from
low energy observations, let us discuss its embedding into the
$SU(2)_L\times SU(2)_R\times SU(4)_c$ model~\cite{ps}. The version of the
model relevant to our discussion is not the one in the original Pati-Salam
paper but rather the one considered in Ref.~\cite{marshak}.
In this model~\cite{marshak}, symmetry breaking from $SU(2)_R\times
SU(4)_c$ to $U(1)_Y\times SU(3)_c$ is implemented by the Higgs fields
belonging to the representation $\Delta_R(1,3,\bar{10})\oplus
\Delta_L(3,1,\bar{10})$
under the $SU(2)_L\times SU(2)_R\times SU(4)_c$ group. Decomposing this
field under the standard model group $SU(2)_L \times U(1)_Y \times SU(3)_c$
gives the various fields in
the model:
\begin{eqnarray}
\Delta_R (1,3,\bar{10}) & \equiv &
\Delta_{u^cu^c}(1,+\frac{8}{3}, 6^*)\oplus
\Delta_{u^cd^c}(1,+\frac{2}{3}, 6^*)\oplus
\Delta_{d^cd^c}(1,-\frac{4}{3}, 6^*)\oplus
\Delta_{u^c\nu^c}(1,+\frac{4}{3}, 3^*)\nonumber\\
&&\oplus
\Delta_{d^c\nu^c}(1,-\frac{2}{3}, 3^*)\oplus
\Delta_{u^ce^c}(1,-\frac{2}{3}, 3^*)\oplus
\Delta_{d^ce^c}(1,-\frac{8}{3}, 3^*)\nonumber\\
&&\oplus
\Delta_{\nu^c\nu^c}(1,0, 1)\oplus
\Delta_{e^c\nu^c}(1,-2, 1)\oplus
\Delta_{e^ce^c}(1,-4, 1)~.
\label{eq:2}
\end{eqnarray}
The Yukawa Lagrangian of this model is given by
\begin{eqnarray}
{\cal L}_I~=~f_{ij}\Psi^{c,T}_iC^{-1}\tau_2\vec{\tau}\cdot
\vec{\Delta}_R\Psi^{c,T}_j~+ (R\leftrightarrow L)+
~H_{ij}^a\Psi_i\Phi_a\Psi^c_j~+~{\rm h.c.}
\label{eq:3}
\end{eqnarray}
where
$\Psi=\pmatrix{u_1 & u_2 & u_3 & \nu \cr d_1 & d_2 & d_3 & e}$.  The $f$ couplings
generate Majorana neutrino masses, while the couplings denoted $H^a$ generate
the Dirac masses for fermions.
Comparing Eq.~(\ref{eq:3}) with Eq.~(\ref{eq:1}), we see that exactly the same interactions
are present in both cases.
 The $S$ field of Eq.~(\ref{eq:1}) is the $\Delta_{\nu^c\nu^c}$
whose vev breaks the gauge group of our model down to the SM gauge
group~\cite{marshak}. We assume that the scale of this symmetry breaking
is anywhere between 1 -- 100 TeV so that both the right--handed neutrinos
as well as the gauge bosons belonging to $SU(4)_c/SU(3)_c$ have masses
around these values.

 We note that while breaking the gauge symmetry only by the $\Delta
(1,3,\bar{10})$ makes the $W_R$ mass scale, $B-L$ breaking scale $v_{BL}$,
and the $SU(4)_c$ breaking scales all equal, and would also
relate the $W_R^\pm$ mass
with the $Z'$ mass. Some of the constraints we derive below require that
$M_{W_R} \gg v_{BL}$. This can be achieved by including a $(1,3,1)$
Higgs field to break the symmetry, which will generate $W_R^\pm$ mass,
but not $Z'$ mass and decouple $W_R^\pm$ mass from $v_{BL}$.  Our results will be valid in the presence of such
$(1,3,1)$ Higgs fields, or in their absence.  In the latter case, the
common scale of $B-L$ symmetry breaking will be required to be
$> 100$ TeV or so.

 The Higgs fields belonging to the $\Delta$ multiplet
will have the following mass pattern:
$ \Delta_{d^c d^c},~  \Delta_{u^c d^c} $ will have mass near a TeV,
whereas
$ \Delta_{u^c u^c} $ will have mass
near 100 TeV.  Such a mass pattern is consistent, since as noted above,  we could
have the $SU(4)_c/SU(3)_c\times U(1)_{B-L}$ boson and the $W_R$ mass
different from the $Z'$ mass and the right--handed neutrino mass scale.

One important point to note is that due to $SU(4)_c$ gauge symmetry, all
three couplings in Eq.~(\ref{eq:1}) become equal to each other i.e.,
$h_{ij}~=~g_{ij}~=~l_{ij}~=~f_{ij}$ of Eq.~(\ref{eq:3}).

In general the neutrino mass in this model is given by a combination of
type I and type II seesaw contributions:
\begin{equation}
 M_\nu=\gamma\frac{v_{wk}^2}{v_{BL}}f-M_\nu^{\rm
Dirac}(v_{BL}f)^{-1}\left(M_\nu^{\rm Dirac}\right)^T~.
\label{eq:4}
\end{equation}
The coupling matrix $f$ that appears in the neutrino mass formula above
is related to the diquark couplings, which lead to FCNC effects,
$n-\bar{n}$ oscillations as well as the baryon asymmetry. They are
therefore very highly constrained.

 In what follows, we assume that $M_\nu^{\rm Dirac}=0$ or very small by
an appropriate choice of the Yukawa couplings of
$\Phi_1 \sim (2,2,1)$ and $\Phi_{15} \sim (2,2,15)$ fields (see Eq.~(\ref{eq:3})). The details
of this are not relevant to the main point of our paper. Since $M_D$
depends on the same parameters as the quark and charged lepton masses, it
is useful to point out that setting $M_D~=~0$ does not lead to any
conflict with realistic fermion mass and mixing patterns. To see this
note that the two bi-doublet fields $\Phi_1$ and $\Phi_{15}$ are both
complex scalars; therefore each field will have
two independent Yukawa couplings with fermions:
\begin{eqnarray}
{\cal L}_{\rm Yukawa} = Y_1 \bar{\psi}_L  \Phi_1 \psi_R + \tilde{Y}_1
\bar{\psi}_L \tilde{\Phi}_1
\psi_R + Y_{15} \bar{\psi}_L \Phi_{15} \psi_R  + \tilde{Y}_{15}
\bar{\psi}_L \tilde{\Phi}_{15}
\psi_R + {\rm h.c.}
\label{eq:5}
\end{eqnarray}
where $\psi_L \sim (2,1,4)$ and $\psi_R \sim (1,2,4^*)$ fermions, and
$\tilde{\Phi}_i \sim \tau_2 \Phi^*_i \tau_2$.
From Eq.~(\ref{eq:5}), it follows that the Dirac mass matrices of the up quark,
down quark, charged lepton, and
the neutrino are all independent.

Once we set $M_\nu^{\rm Dirac}=0$, we
can directly
link the neutrino mass matrix to the coupling matrix $f$. The advantage
of this is that the requirement of adequate baryogenesis as well as consistency with
FCNC and other constraints fix not only the neutrino
mass matrix, but also the mass spectrum of the theory. The FCNC
constraints come from the fact that $ \Delta_{u^c u^c},~\Delta_{d^c d^c},~\Delta_{u^c d^c}$
fields have masses in the multi-TeV range and can lead to sizable
$K^0-\bar{K}^0,~ D^0-\bar{D}^0$ and $B^0_{d,s}-\bar{B}^0_{d,s}$ mixings.
They in turn
 severely constrain the pattern of the Yukawa couplings $f_{ij}$ and thereby
the neutrino mass matrix.

\section{Low energy constraints on the model}
In this section, we discuss the tree level
flavor changing neutral current contributions to processes such as
$K-\bar{K}$, $B_{d,s}-\bar{B}_{d,s}$, $D-\bar{D}$ mixings from the
di-quark Higgs field exchanges. We have to
make sure that they are not in conflict with observations. One cannot make
the di-quark scalar masses very large to satisfy the FCNC constraints, since
successful post-sphaleron baryogenesis requires the masses of at least two of these scalars
to be not more than about a TeV. Similarly, the
doubly charged scalar bosons from the same multiplet will contribute to rare
processes such as $\mu\to 3 e$ via tree level diagrams.
Neutrino oscillation data, on the other
hand, suggest specific form of the $f$ matrix.  We need to examine if
these dual requirements can be simultaneously met.  We have found that
indeed this can
be satisfied, but only for an inverted mass hierarchy spectrum for the
neutrinos.

 To discuss the constraints
on the couplings $f_{ij}$ and masses of $ \Delta_{u^cu^c},~\Delta_{d^cd^c},~\Delta_{u^cd^c}$
implied by these
considerations, we first note that above the $SU(4)_c$ scale, all
couplings to diquarks and dileptons are given by a single matrix
$f_{ij}$. The form of
this matrix can be specified in any basis without loss of generality and
we specify them in the basis in which the down quarks are mass
eigenstates. In this basis, the $f_{ij}$ couplings split up into the
following depending on which quarks they couple to:
$ f_{dd},~f_{ud}$ and $f_{uu}$, where $f_{dd}$ indicates the coupling to
$d^c d^c$, etc.  Assuming for simplicity that $CP$ is not broken by the
vacuum expectation
values of the bidoublet fields (so that the left--handed and right--handed
CKM matrices are equal to each other), we get
\begin{eqnarray}
f_{ud}&=&U_{CKM}f_{dd} \nonumber\\
f_{uu}&=&U_{CKM}f_{dd}U^T_{CKM}\nonumber\\
f_{\nu\nu}&=&U_l f_{dd} U^T_l=f_{ee}
\label{eq:6}
\end{eqnarray}
where $U_{CKM}$ is the quark rotation matrix and $U_l$ is the matrix that
makes the charged leptons diagonal. Clearly, it is $f_{\nu\nu}$ which
determines the neutrino mass matrix in the type II seesaw case.

 In this basis, first there are constraints from
flavor changing processes such as $K-\bar{K}$, $B_{s,d}-\bar{B}_{s,d}$ and
$D-\bar{D}$ mixings. Below we list the constraints~\cite{PDG} and their
implications for the parameters of the model:
\begin{eqnarray}
K^0(d\bar{s})-\bar{K^0}(\bar{d}s)&:&
\frac{f_{dd,11}f_{dd,22}}{[m_{\Delta_{d^cd^c}}/{\rm TeV}]^2}\lsim
3.3\times 10^{-6}
\label{eq:7}\\
B_s^0 (s\bar b)-\bar{B_s^0} (\bar s b)&:&
 \frac{f_{dd,22}f_{dd,33}}{[m_{\Delta_{d^cd^c}}/{\rm TeV}]^2}\lsim
2.0\times 10^{-4}
\label{eq:8}\\
B_d^0 (d\bar b)-\bar{B_d^0} (\bar d b)&:&
 \frac{f_{dd,11}f_{dd,33}}{[m_{\Delta_{d^cd^c}}/{\rm TeV}]^2}\lsim
7.6\times 10^{-6}
\label{eq:9}\\
D^0 (u\bar c)-\bar{D^0} (\bar u c)&:&
\frac{f_{uu,11}f_{uu,22}}{[m_{\Delta_{u^cu^c}}/{\rm TeV}]^2}\lsim
2\times 10^{-6}
\label{eq:10}
\end{eqnarray}

In addition, lepton family number
violating modes~\cite{mace} such as $\mu\to 3e$ imply
 \begin{equation}
 \frac{f_{ee,11}f_{ee,12}}{[m_{\Delta^{++}}/{\rm
TeV}]^2}\lsim 3.3\times 10^{-5}~.
\label{eq:11}
\end{equation}
This can be satisfied by requiring the $\Delta^{++}$ mass to be in the
100 TeV range for our
choice of $f_{12,11}$ as we see below. The constraints from the various
$\tau$ decay modes can then be easily satisfied for this limit on the
$\Delta^{++}$ mass and we do not give those constraints here.

Another constraint on the parameters of the theory comes from the
 present limits on $n-\bar{n}$ oscillation period.
$\tau_{n-\bar{n}}\geq 10^8$ sec. \cite{milla,kajita} implies that the
strength
$G_{n-\bar{n}}$ of the $\Delta B= 2$
transition is $\leq 10^{-28}$ GeV$^{-5}$. In a generic model of this
type, $n-\bar{n}$ oscillations arise from the tree diagram in
Fig.~\ref{fig:tree} (see Sec. 5) and we find that
\begin{eqnarray}
G_{n-\bar{n}}\simeq \frac{\lambda_1 \left\langle S \right \rangle
f^2_{dd,11} f_{uu,11}}{M_{\Delta_{d^cd^c}}^4
M_{\Delta_{u^cu^c}}^2}~+~\frac{\lambda_2
\left\langle S
\right \rangle
f_{dd,11}f^2_{ud,11}}{M_{\Delta_{d^cd^c}}^2M_{\Delta_{u^cd^c}}^4}\leq
10^{-28}~ {\rm
GeV}^{-5}~.
\label{eq:12}
\end{eqnarray}
We discuss this further in our model in Sec. 6.

The nontrivial aspect of this model is that the same set of parameters
responsible for baryogenesis is directly related to neutrino masses
and mixings and must be such that they satisfy the strong FCNC
constraints listed above. Note that we cannot suppress the FCNC effects
by simply raising the masses of $\Delta_{d^cd^c},~\Delta_{u^cd^c}$ particles
since in that case we
cannot satisfy the desired constraints for adequate baryogenesis.

\section{Inverted neutrino mass hierarchy from the FCNC constraints}
In this section, we address the question of how we satisfy these
constraints and yet obtain concordance with neutrino oscillation
observations. It turns out that if we choose $f_{dd}$ matrix as (in a basis
where down quarks are mass eigenstates)
\begin{equation}
f_{dd}=\left(\begin{array}{ccc}
0&0.95&1\\
0.95&0&0.01\\
1&0.01&-0.0627357\end{array}\right),
\label{eq:13}
\end{equation}
then for TeV scale $\Delta_{d^cd^c}$ and $\Delta_{u^cd^c}$ and 100 TeV mass
for $\Delta_{u^cu^c}$, we can satisfy
all the hadronic constraints. Such a choice will automatically satisfy
$K-\bar{K}$ and $B_{d,s}-\bar{B}_{d,s}$ mixing constraints,
owing to the zeros in the diagonal entries.
In the leptonic sector, as already noted,
the most stringent
constraint comes from $\mu\to 3e$ and it requires that
$\Delta^{++}$ mass also be of order 100 TeV or so.

\begin{figure}[htb]
\centering
\includegraphics[width=2 in]{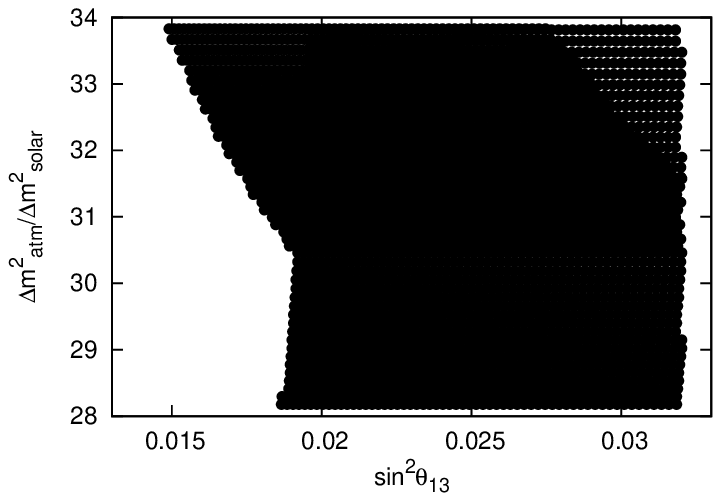}
\includegraphics[width=2 in]{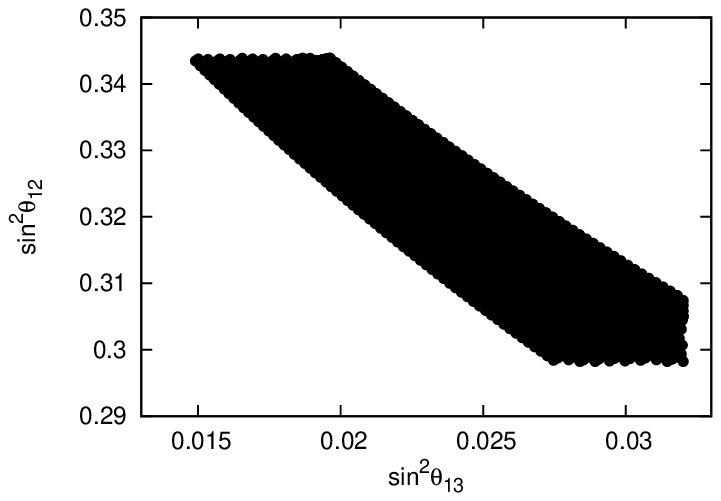}
\includegraphics[width=2 in]{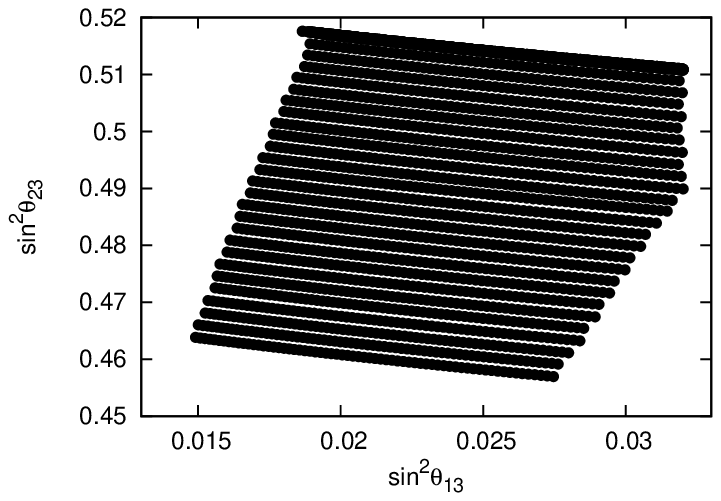}
\caption{We give the predictions for neutrino oscillation parameters for
the allowed ranges of the diquark scalar couplings in our model. Note
the lower limit on the $\theta_{13}$ of about $0.1$.} \label{scatter}
\label{fig:scatter}
\end{figure}

Then we use the following unitary transformation to rotate $f_{dd}$ to get
the neutrino mass matrix:
\begin{equation}
U_l=\left(\begin{array}{ccc}
\cos\Theta&\sin\Theta&0\\
-\sin\Theta&\cos\Theta&0\\
0&0&1
\end{array}\right)
\label{eq:14}
\end{equation}
with $\Theta=0.23$. This gives
\begin{equation}
f_\nu=U_l f_{dd} U_l^T=
\left(\begin{array}{ccc}
0.421751 & 0.85125 & 0.975946\\
0.85125 & -0.421751 & -0.218241\\
0.975946 & -0.218241 & -0.0627357
\end{array}\right)
\label{eq:15}
\end{equation}
This matrix must be multiplied by the $SU(2)_L$ triplet vev $v_L$ (which is
much smaller than the electroweak vev in Type II seesaw) to give the
neutrino mass matrix. For $v_L=0.03 $ eV, we get the three neutrino
masses to be
\[m_1=0.0410~{\rm eV},~m_2=-0.0417~{\rm eV},~m_3=-0.0012~{\rm eV}\]
which yields the correct solar-to-atmospheric neutrino mass
square difference $\Delta m^2_{\rm atm}/\Delta m^2_{\rm solar}\simeq 30$
and the PMNS mixing angles
$\theta_{12}=35.6^0,~\theta_{23}=46^0$ and $\theta_{13}=8^0$.
This value of $\theta_{13}$ is observable in the ongoing
(Double CHOOZ)~\cite{dc} and planned (Daya Bay)~\cite{db} experiments
and should provide a test of the model.
Secondly we predict that the neutrinoless double $\beta$-decay experiments
should observe a Majorana neutrino mass at the 10-20 meV level which is
perhaps within reach of the next round of neutrinoless double $\beta$-decay
experiments. In Fig.~\ref{scatter}, we present two scatter plots that
display the preference of oscillation parameters in our model.
We clearly see the lower bound on the $\theta_{13}$ from them.

We will see below that this form of the $f$ matrix satisfies the baryon
asymmetry  constraints as well as the $n-\bar{n}$ constraints.

\section{Origin of matter}
Before proceeding to the discussion of how baryon asymmetry arises
in this model, let us first sketch the cosmological sequence of events
starting at the $SU(4)_c$ scale that leads up to this. For
temperatures above the $SU(4)_c$ scale of about
100 TeV, there is no $B-L$ violation. The sphalerons are active and
therefore erase any pre-existing $B+L$ asymmetry in the universe. So if
there was a primordial GUT scale generated baryon asymmetry that
conserved $B-L$ (like that in most $SU(5)$ and some $SO(10)$ models), it will
be erased by sphalerons. Any baryon asymmetry residing in $B-L$ violating
interactions will however survive.

 Below the $SU(4)_c$ scale, $B-L$ violating interactions arise e.g. $SS\to
e^+e^-$, and will be in equilibrium together with the $\Delta B=2$
interactions. So together they will erase any pre-existing baryon or
lepton asymmetry. Thus in models of this kind, baryon asymmetry of the
universe must be generated fresh below the sphaleron decoupling
temperature.

In order to sketch how fresh baryon asymmetry arises in our model, we
assume
the following mass hierarchy between the $S_r$ field and the
$\Delta_{d^cd^c},~\Delta_{u^cu^c},~\Delta_{u^cd^c}$ fields:
\[m_t < M_S(\sim 500~ {\rm GeV}) <  M_{\Delta_{d^cd^c}}\sim M_{\Delta_{u^cd^c}}(\sim 1~{\rm~
TeV})\ll M_{\Delta_{u^cu^c}}(\sim 100~ {\rm TeV}),\]
where $m_t$ is the top quark mass.

 Between $1 \leq T\leq 100$ TeV, the $\Delta B=2$
interaction rates go like
\begin{equation}
\Gamma (\Delta B=2)\sim\frac{f^6_{11}}{(2\pi)^9}T
\label{eq:16}
\end{equation}
and are therefore in equilibrium if some of the $f_{ij}$'s are above
$0.3$ as in our case.

Below $T\sim 1$ TeV, the $\Delta B=2$ processes
such as the decay $S_r\to 6q^c+6\bar{q}^c,
~(\bar{q}^c,q^c)+~S_r\to 5 (q^c, \bar{q}^c)$ occur at a rate given by
\begin{eqnarray}
\Gamma (\Delta B=2)\sim\frac{100 f^6_{ud,12}}{(2\pi)^9}\frac{T^{13}}{(6
M)^{12}}
\label{eq:17}
\end{eqnarray}
where $M\sim $ TeV, the average mass of the $\Delta_{d^cd^c},~\Delta_{u^cd^c}$ particles
which are still in equilibrium. The $\Delta_{u^cu^c}$ is about 100 TeV
and hence its contribution to these processes is
more suppressed compared to that of $\Delta_{d^cd^c},~\Delta_{u^cd^c}$. This decay
then goes out of
equilibrium somewhat below the TeV temperature range. One impact of this
is that these interactions being in equilibrium above $T\sim $ TeV
erase any pre-existing baryon asymmetry as discussed above.

By the time the universe cools to temperature near or slightly below
$M_S$, its decay channels can start if the rates are faster compared to
the Hubble expansion rate. Let us therefore estimate the various decay
rates:

There are four decay modes which are competitive with each other:
(i) $S_r\to 6 q^c$; (ii) $S_r\to Zf^c\bar{f}^c$; (iii) $S_r\to ZZ$ and
(iv) $S_r\to \tau e$.\footnote{The $S_r\to W^+W^-$ is suppressed by $W_L-W_R$
mixing parameter which can be adjusted to be small.} We discuss
them below.
\begin{figure}[htb]
\psfrag{S}{\tiny {$S_r$}}
\psfrag{X}{\tiny{$\Delta_{d^cd^c}$}}
\psfrag{Y(Z)}{\tiny{$\Delta_{u^cu^c(u^cd^c)}$}}
\psfrag{X(Z)}{\tiny{$\Delta_{d^cd^c(u^cd^c)}$}}
\psfrag{dc}{\tiny{$d^c$}}
\psfrag{uc(dc)}{\tiny{$u^c(d^c)$}}
\psfrag{uc}{\tiny{$u^c$}}
\psfrag{dc(uc)}{\tiny{$d^c(u^c)$}}
\centering \epsfysize=2.0in \epsffile{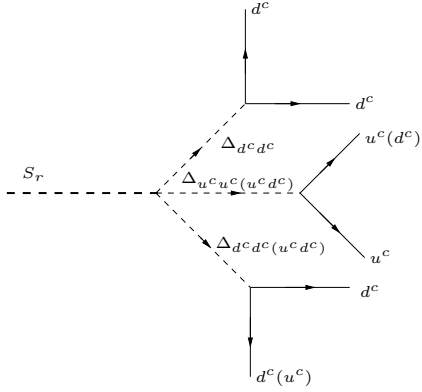}
\caption{Tree
level diagrams contributing to $S_r$ decays into 6 anti-quarks.
There are other diagrams where $S_r$ decays into 6 quarks,
obtained from the above by reversing the arrows of the quark
fields.}
\label{fig:tree}
\end{figure}

(i) {\underline \bf $S_r\to 6 q^c$}~: The diagram for this is given in
Fig.~\ref{fig:tree}. Since $M_S \gg m_t$, in its decay
all modes will participate. Including all the modes, we find the decay
rate to be:
\begin{eqnarray}
\Gamma(S_r\to 6q^c)~\simeq \frac{36}{(2\pi)^9}\frac{({\rm Tr}[f^\dagger f])^3\lambda^2
M^{13}_S}{ (6M_\Delta)^{12}}
\label{eq:18}
\end{eqnarray}
where we have chosen $\lambda_1=\lambda_2\equiv \lambda \sim 0.1$.
Taking as an example a typical set of parameters $M_\Delta \simeq 2
M_S\sim 1$ TeV and taking the parameters for the $f$ matrix elements from
Eq.~(\ref{eq:13}), we get $\Gamma(S_r\to 6 q^c)\sim 7.5\times 10^{-17}$ GeV.

(ii) {\underline \bf $S_r\to Z+f^c\bar{f}^c$}~: This arises from the $S\bar{S}$
coupling to $Z'Z'$ with one of the $Z$'s mixing with $Z'$ (Fig. 3)
and the virtual $Z'$ decaying to $f^c\bar{f}^c$. This occurs only for $T\leq
v_{wk}$. This is because for $T\geq
v_{wk}$, $Z-Z'$ mixing disappears. Below the electroweak symmetry
breaking temperature, this mixing denoted below by $g_{ZZ'}$ becomes
effective and is given by
\begin{eqnarray}
 g_{ZZ'}=\frac{g^2\cos^2\theta_W}{\sqrt{\cos{2\theta_W}}}
\left(\frac{M_Z}{M_{Z'}}\right)^2v_{BL}
\label{eq:19}
\end{eqnarray}
which leads to the new $S_r$ decay mode (Fig. 3) given above.
\begin{figure}[htb]
\begin{center}
\psfrag{fc}{$\bar{f^c}$}
\psfrag{f}{$f^c$}
\psfrag{S}{$S_r$}
\psfrag{Z}{$Z$}
\psfrag{Z'}{$Z'$}
\includegraphics[width=5cm]{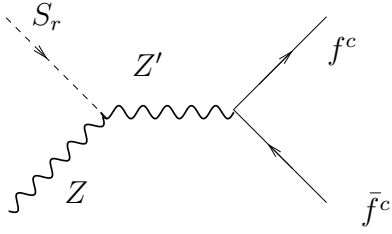}
\caption{Feynman diagram for $S_r$ decay to $Zf^c\bar{f^c}$ vis $Z-Z'$
mixing}
\end{center}
\label{fig:sz}
\end{figure}
This decay rate is given by
\begin{eqnarray}
 \Gamma(S_r\to Zf^c\bar{f^c}) \simeq
\frac{7.0\times 10^{-2}~{\rm GeV}^2}{M_sM_{Z'}^6}\left
[M_s\sqrt{M_s^2-M_Z^2}~\left(6M_s^4-19M_s^2M_Z^2+28M_Z^4\right)
\right.\nonumber\\
~~~\left.-3M_Z^4(M_s^2+4M_Z^2)\log
\left(\frac{M_s+\sqrt{M_s^2-M_Z^2}}{M_Z}\right)\right]~.
\label{eq:20}
\end{eqnarray}
For our choice of parameters and $M_{Z'}\sim 100$ TeV, we find that
$\Gamma (S_r\to Z f^c\bar{f}^c)\simeq 6.6\times 10^{-18}$ GeV and is therefore
slower than the $6q^c$ decay rate.

(iii) {\underline \bf $S_r\to ZZ$}~: This decay mode arises from $Z-Z'$
mixing with the decay width given by
\begin{equation}
\Gamma(S_r\to ZZ)=\frac{g_{ZZ}^2M_S^3}{128\pi M_Z^4}
\left(1-\frac{4M_Z^2}{M_S^2}\right)^{1/2}
\left[1-\frac{4M_Z^2}{M_S^2}+\frac{12M_Z^4}{M_S^4}\right],
\label{eq:21}
\end{equation}
where the $S_rZZ$ vertex is given by
\begin{equation}
g_{ZZ}=\frac{1}{2}g^2\cos^2\theta_W v_{BL}\left(\frac{M_Z}{M_{Z'}}\right)^4~.
\label{eq:22}
\end{equation}
For $M_{Z'}\sim 100$ TeV and $M_S=500$ GeV, we get $\Gamma(S_r\to ZZ)\simeq
2.2\times 10^{-19}$ GeV which is much smaller than the $6q^c$ decay rate.

\begin{figure}[htb]
\begin{center}
\includegraphics[width=10cm]{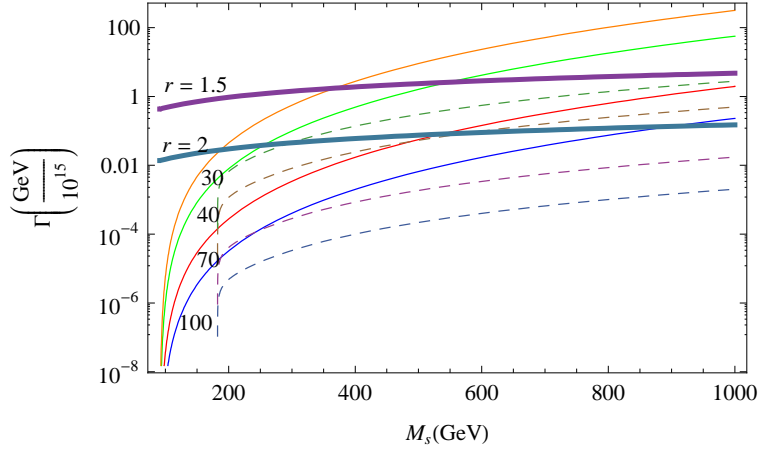}
\end{center}
\caption{The $S_r\to Zf^c\bar{f^c}$ (thin solid lines) and $S_r\to ZZ$ (thin
dashed lines) decay rates for various values of $v_{BL}$ (in TeV).
The thick solid lines correspond to the $S_r\to 6q^c$ decay rates for two
typical values of $r=M_{\Delta_{u^cd^c,d^cd^c}}/M_S$. We see that for
$v_{BL} \geq 40$ TeV, the six quark decay mode dominates for a large range of
$M_S$.}
\label{fig:decaywidth}
\end{figure}

In Fig.~\ref{fig:decaywidth}, these decay rates are plotted against the mass of the scalar field
for various values of $v_{BL}$.
Note that $M_{Z'}$ is related to $v_{BL}$ as follows:
\[M_{Z'}^2\simeq \frac{2g^2v_{BL}^2\cos^2\theta_W}{\cos{2\theta_W}}\]

\begin{figure}[h!]
\psfrag{S}{\tiny{$S_r$}}
\psfrag{WR}{\tiny{$W_R$}}
\psfrag{eR}{\tiny{$e^-_R$}}
\psfrag{tauR}{\tiny{$\tau^+_R$}}
\psfrag{+}{+}
\psfrag{N1}{\tiny{$N_{R,1}$}}
\psfrag{N3}{\tiny{$N_{R,3}$}}
\centering \epsfysize=2.0in \epsffile{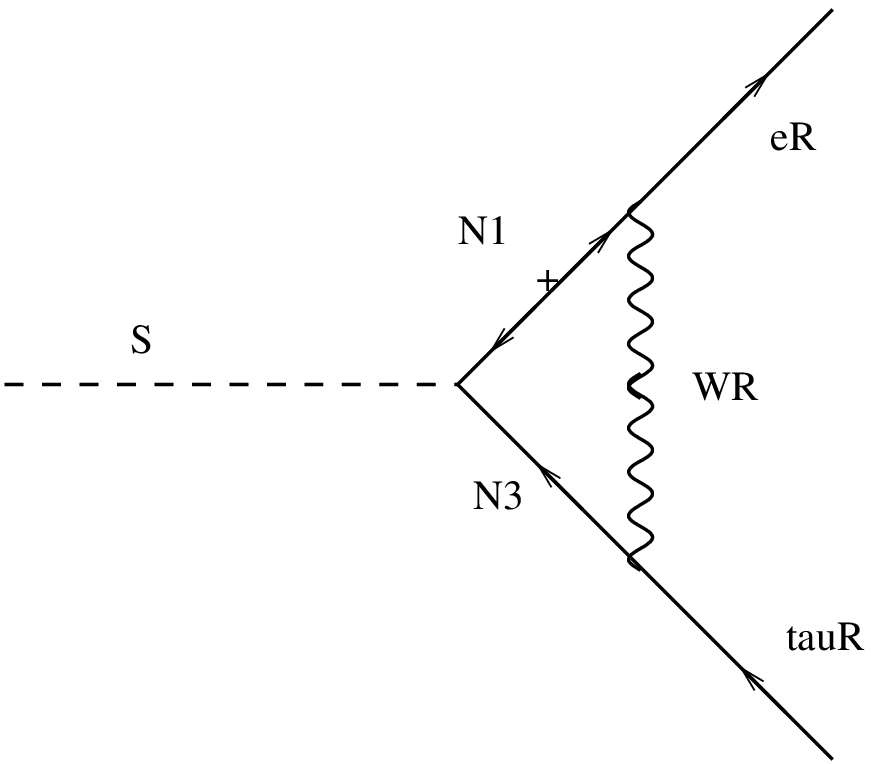}
\caption{$S_r\to e\tau$ decay}
\label{fig:tau}
\end{figure}

(iv) {\underline \bf $S_r\to \tau + e$}: This decay mode arises
from the Feynman diagram in Fig.~\ref{fig:tau} and its rate can be estimated to be:
\begin{eqnarray}
\Gamma (S_r\to \tau+e) ~\simeq \frac{f_{13}^2 g^4 (m_\tau M_{\nu_R})^2 M_S}
{12\pi
(16\pi^2)^2 32 M^4_{W_R}}~.
\label{eq:23}
\end{eqnarray}
This width is estimated to be $\Gamma (S_r\to \tau+e)\simeq 9\times 10^{-20}$
GeV. Therefore this is also much smaller than the decay rate to six quark
modes.

At the time the universe has a temperature of $\sim M_S$ or slightly
below so that it is out of equilibrium from the cosmic soup, the Hubble
expansion rate is $\sim \sqrt{g_{*}}M^2_S/M_{Pl}\sim 2.5 \times 10^{-12}$
GeV implying that all the above decay modes are out of equilibrium. Since
the decay rate remains constant below $T\sim M_S$, but the expansion rate
of the universe is slowing down as it expands, there will come a time (or
temperature $T_d$) when the dominant decay $\Gamma (S_r\to 6q)\simeq
H(T_d)$. At that point the $S_r$ particle will start decaying and produce
the baryon asymmetry as in Ref. \cite{bmn1}.

At this temperature
which is far below the masses of the $\Delta_{u^cd^c},~\Delta_{d^cd^c}$ particles, the
decay processes $\Delta_{q^cq^c}\rightarrow
q^c q^c$ being very fast have depleted all the diquark Higgses and
have left
only the $S_r$ particles to survive along with the usual standard model
particles. The primary decay modes of $S_r$ are $S_r\rightarrow
u^cd^cd^cu^cd^cd^c$ and $S_r \rightarrow \bar{u}^c
\bar{d}^c\bar{d}^c\bar{u}^c \bar{d}^c\bar{d}^c$ as already noted
(Fig.~\ref{fig:tree}). Other decay modes are negligible as discussed.

We have to make sure that the decay of $S_r$ starts below the sphaleron
decoupling temperature and above the QCD phase transition temperature. To
check if this indeed happens in our model, let us calculate the $T_d$:
\begin{eqnarray}
T_d &\simeq& \left[\frac{36 \lambda^2 ({\rm Tr}[f^\dagger f])^3 M_{\rm
Pl}M_S^{13}}{(2\pi)^9
1.66 g_*^{1/2}(6M_\Delta)^{12}}\right]^{1/2}\nonumber\\
&\simeq& 6.1~{\rm
GeV}^{1/2}\left(\frac{M_S^{13}}{M_{\Delta}^{12}}\right)^{1/2}
\label{eq:24}
\end{eqnarray}
For $M_S\sim 500$ GeV and
$M_{\Delta} \sim M_{\Delta_{u^c d^c}} \sim M_{\Delta_{d^cd^c}}\sim 1$ TeV, we get
$T_d\simeq 2$ GeV which is comfortably above the QCD phase
transition temperature.

It is worth emphasizing that if we increased the sextet scalar masses
arbitrarily to satisfy the FCNC constraints, this will lower the $T_d$ to
undesirable values below the QCD temperature. One may think that we could
simultaneously increase the value of $M_S$ but as we will see below, the
magnitude of the baryon asymmetry goes inversely like the square of $M_S$ and
increasing it above 500-600 GeV will suppress the baryon asymmetry to a
level below the observations.
\begin{figure}[htb]
\psfrag{S}{\tiny{$S_r$}}
\psfrag{X}{\tiny{$\Delta_{d^cd^c}$}}
\psfrag{Z}{\tiny{$\Delta_{u^cd^c}$}}
\psfrag{W}{\tiny{$W_L$}}
\psfrag{+}{\tiny{$+$}}
\psfrag{dic}{\tiny{$d_i^c$}}
\psfrag{djc}{\tiny{$d_j^c$}}
\psfrag{dk}{\tiny{$d_k$}}
\psfrag{dkc}{\tiny{$d_k^c$}}
\psfrag{um}{\tiny{$u_m$}}
\psfrag{umc}{\tiny{$u_m^c$}}
\psfrag{ua}{\tiny{$u_\alpha$}}
\psfrag{uac}{\tiny{$u_\alpha^c$}}
\psfrag{db}{\tiny{$d_\beta$}}
\psfrag{dbc}{\tiny{$d_\beta^c$}}
\psfrag{dlc}{\tiny{$d_l^c$}}
\psfrag{unc}{\tiny{$u_n^c$}}
\centering \epsfysize=2.0in \epsffile{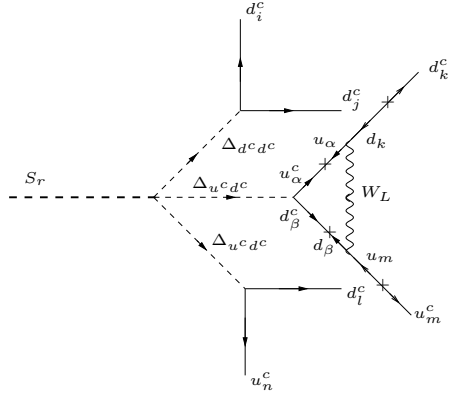}
\caption{One loop
vertex correction diagram for the $B$-violating decay
$S_r\rightarrow 6q^c$. There are also wave function corrections involving the exchange of $W^\pm$ gauge bosons, which
are somewhat smaller.}
\label{fig:loop}
\end{figure}

The calculation of the baryon asymmetry is same as in Ref. \cite{bmn1} and
we do not repeat it here except to give a brief summary for the
particular $f$ texture in our model. First we note that since only
$f_{31}$ and $f_{21}$ are the dominant contributions, there is a flavor
factor of $64\equiv ({\rm Tr}[f^\dagger f])^3$ in the absolute decay width. Next
we calculate the baryon asymmetry from the vertex correction via the $W$
boson exchange (Fig. 6), which dominates the baryon asymmetry. To do this
we note that the dominant contribution comes from making the $f$ matrix
complex (e.g. $f_{33}$ with a maximal complex phase and all other
parameters real). This of course does not affect the neutrino fit
discussed earlier. In terms of the flavor combinations that give the
dominant contributions, they come from products like $f^2_{31}f_{33}$ of
which there are six combinations. This gives
 \begin{equation}
{\epsilon_B^{\rm vertex} \over {\rm Br}} \simeq - {\alpha_2 \over
4} {6~{\rm Im}~[ f^2_{31}m_t V_{tb} m_b f^*_{33} m_t V_{tb}{m}_b] \over
({\rm Tr}[f^\dagger f])^3 M^2_WM^2_S }~.
\end{equation}
We have also assumed that $M_{S} \gg m_t$. Note that if we increased
$M_S$ above 500 GeV or so, the generated baryon asymmetry will fall short
of the observed values.

This gives for the baryon asymmetry at $T=T_d$:
$\epsilon_B\sim (2-3) \times 10^{-8}$.
To compare it with the observed $\eta_B$, we divide this by $g_*(200~
{\rm MeV})/g_*(1~{\rm eV})\sim 62.75/5.5 = 11.4$ and apply an additional dilution
factor of 0.25 (see discussion below) which gives us the desired value.
Note that the observed value of the asymmetry is
$\eta^{\rm CMB}_B\sim 6\times 10^{-10}$~\cite{WMAP3}.

Since the $S_r$-particle decays far below its mass to generate the baryon
asymmetry, we have to take into account the effect of its decay, which
as we explain below amounts to a dilution of the original baryon
asymmetry calculated. In order to estimate the dilution factor, we note
that $S_r$-decay will release all the energy in its mass to lighter
relativistic particles which will thermalize with the rest of the cosmic
fluid and in the process raise its temperature which will increase the
entropy and hence dilute the net baryon asymmetry. Suppose the decay
temperature is $T_d$. Energy conservation then gives
\begin{eqnarray}
&&\left.\rho_S+\rho_{\rm rel}\right|_{T_d}
\simeq \left.\rho_{\rm rel}\right|_{T_>}, \nonumber\\
&{\rm or},&
\frac{1.2}{\pi^2}T_d^3 M_S+\frac{\pi^2}{30}g_*T^4_d~=
\frac{\pi^2}{30}g_*T^4_>~.
\label{eq:26}
\end{eqnarray}
Solving this one finds for the dilution factor $d$ that
\begin{eqnarray}
d~\equiv \frac{T^3_d}{T^3_>}~\simeq \frac{0.32g_*T_>}{0.12 M_S+0.32
g_*T_d}~.
\label{eq:27}
\end{eqnarray}
For $M_S\sim 500$ GeV and $T_d\simeq 1$ GeV, $d\simeq 0.25$.

There is another factor coming from the fact that the baryon asymmetry
generated is at $T_d\sim 1$ GeV where $g_*= 62.75$ whereas the
measured value is at $T_{\rm rec}$ where $g_* = 5.5$.  All these dilution factors
have been taken into account in our estimate of final baryon asymmetry, which
is in agreement with observations.

\section{Prediction of observable neutron-antineutron oscillation time}
In this section, we discuss the prediction of the model for
neutron-anti-neutron oscillation. In order to estimate the
$n-\bar{n}$ oscillation, let us first recall that the only
contribution to this process comes from the right--handed sector
since the vev of the $\Delta_{\nu^c\nu^c}$ is in the 100 TeV range
and that of its left-handed counterpart is in the eV range.

\begin{figure}[htb]
\psfrag{S}{\tiny{$S_r$}}
\psfrag{uc}{\tiny{$u^c$}}
\psfrag{dc}{\tiny{$d^c$}}
\psfrag{bc}{\tiny{$b^c$}}
\psfrag{tc}{\tiny{$t$}}
\psfrag{W}{\tiny{$W_L$}}
\psfrag{ud}{\tiny{$\Delta_{u^cd^c}$}}
\psfrag{dd}{\tiny{$\Delta_{d^cd^c}$}}
\centering \epsfysize=2.0in \epsffile{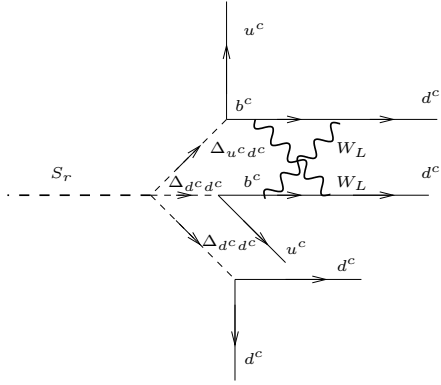}
\caption{Loop diagram for $n-\bar{n}$ oscillation}
\label{fig:nnbar}
\end{figure}

There are two types of contributions to $n-\bar{n}$ oscillation from the
right--handed sector: the one involving two $d^cd^c$ type
and one $u^cu^c$ type bosons of the right--handed sector and another which
involves two $u^cd^c$ and one $d^cd^c$ type $\Delta$ boson.
Since the diagonal $11$ and $22$ entries of $f_{dd}$ are close to zero,
the first contribution is actually much smaller than the second one.
The second type generates an effective operator of the
form $u^cd^cb^cu^cd^cb^c$. To get
$n-\bar{n}$ oscillation, we will have to change the two $b^c$ quarks to two
$d^c$ quarks by second order  weak interactions (see Fig.~\ref{fig:nnbar}).

From Fig.~\ref{fig:nnbar}, we see that the six quark operator for $n-\bar{n}$
oscillation has $\gamma_\mu$'s in the Lorentz structure i.e.
$(\bar{u_R}\gamma_\mu \bar{d_L}^T )^2d^cd^c$. This is different form for
the $\Delta B=2$ operator whose matrix element has not been evaluated in
the literature. For the strength of this operator we estimate
\begin{eqnarray}
G_{n-\bar{n}}~\simeq \frac{f_{ud, 11}f_{ud,13}f_{dd,13}\lambda
v_{BL}}{M^4_{\Delta_{u^cd^c}}M^2_{\Delta_{d^cd^c}}}\frac{g^4 V^2_{td} m^2_b
m^2_t}{(16\pi^2)^2m^4_W} {\rm log}({m_b^2 \over m_W^2})
\end{eqnarray}
 This gives $G_{n-\bar{n}}\sim 10^{-30}$ GeV$^{-5}$. Taking the hadronic
``dressing'' of quark to hadrons to be a factor of $10^{-4}$, we estimate
$n-\bar{n}$ transition time of $10^{9-10}$ sec. given the uncertainties
in the parameters. The present lower limit on this transition time is
$10^8$ sec. from the Grenoble experiment~\cite{milla} as well as from nuclear
decay experiments~\cite{kajita}. Our predicted value
is accessible to current experiments
under discussion at DUSEL as well as other facilities~\cite{yuri}.

\section{Other implications and tests of the model}

\noindent(i) As noted in Sec. 2, a crucial prediction of our quark-lepton
unified model of
post-sphaleron baryogenesis is that neutrinos must be Majorana fermions
and exhibit an inverted mass hierarchy form with large value for $\theta_{13}$.
This should be testable in long base line experiments as well as the
ongoing and planned reactor experiments searching for $\theta_{13}$ and
neutrinoless double beta decay searches~\cite{review}. It
is perhaps worth noting that there is indication of a non-zero
$\theta_{13}$ from already existing neutrino oscillation
data~\cite{fogli}.

\noindent(ii) Our theory is also testable in collider experiments such as
the LHC since we have colored diquark scalar fields with masses in
the TeV range. It is clear from the form of the $f_{ud}$ matrix that in a
$p{p}$ collision, the valence quarks in the two protons could produce
the $\Delta_{u^c d^c}$ field which could then decay to $t+$ jets. This
could either be an $s$-channel single production~\cite{yu} or Drell-Yan pair
production~\cite{wang}.  The $s$-channel
process will have a resonant enhancement which can give a signal above
the standard model background. The Drell-Yan pair production could give
signals of type $bbl^{\pm}l^{\pm}jj+$ missing $E_T$.
 Unlike the $s$-channel process, the Drell-Yan pair
production has the advantage of not being dependent on
the specific flavor texture of the two quark couplings $f$ and is
 promising for color sextet masses upto a TeV~\cite{wang}. It would
therefore be important to search for these particles at LHC. Their discovery
will signal a completely different direction for unification beyond the
standard model than the conventional SUSY--GUT theories.

\noindent(iii) In our model, since there is a mass hierarchy between the
$\Delta_{u^cd^c},~\Delta_{d^cd^c}$ and $\Delta_{u^cu^c}$ masses i.e. $M_{\Delta_{u^cd^c}},~M_{\Delta_{d^cd^c}}\ll
M_{\Delta_{u^cu^c}}$, a one loop level box graph induced by the trilinear
coupling $\lambda v_{BL} \Delta_{u^cd^c}\Delta_{u^cd^c} \Delta_{d^c d^c}$ will induce a quartic coupling
$(\Delta^\dagger_{u^cd^c}\Delta_{u^cd^c})^2$ coupling with a strength
$\lambda_{eff}\simeq -\frac{1}{16\pi^2}\left(\frac{\lambda
v_{BL}}{M_{\Delta_{d^cd^c}}}\right)^4$. This can lead to color breaking
unless  $\lambda_{eff}\leq 1-2$. This can be satisfied by lowering the
$v_{BL}$ scale to about 50 TeV with $\lambda \sim 0.1$. In order to
reconcile this lower value with constraints from $\mu\to 3e$, we can
introduce a multiplet of type $(1,3,1)$ with a vev in the 100 TeV
range which gives mass to the $W_R$ and the $\Delta^{++}$. This vev
decouples the $B-L$ breaking scale $v_{BL}$ from the
masses of the $W_R$ and $\Delta^{++}$ fields. In this case, one can
keep the $B-L$ breaking scale near 50 TeV while keeping the $W_R$ and
$\Delta^{++}$ mass around 100 TeV as required by the $\mu\to 3e$ and
$S_r\to \tau+e$ constraints.

\noindent(iv) Finally, note that a priori in the model there could be a
coupling
of type $\Delta^\dagger \Delta {\rm Tr} [\Phi^\dagger \Phi]$, which will 
induce a $S_r$ decay to two SM Higgs fields. We assume that this
parameter is very small . This assumption could be justified in
supersymmetric
extensions of the model where such terms are forbidden by holomorphy of
the superpotential.

 \section{Conclusion}
In summary, we have pointed out that the post-sphaleron baryogenesis
mechanism proposed in Ref.~\cite{bmn1,bmn2} can be naturally embedded
into a 100 TeV scale quark-lepton unified $SU(2)_L\times SU(2)_R\times
SU(4)_c$ model. If we further assume that the neutrino masses in this
model arise via the type II seesaw mechanism, then the couplings
responsible for baryogenesis and neutrino masses get intimately linked to
one another. In this case adequate baryogenesis predicts that neutrino
mass ordering must be inverted with large $\theta_{13}$, a prediction
that can be tested in ongoing neutrino experiments.

The work of KSB is supported by DOE grant Nos. DE-FG02-04ER46140
and DE-FG02-04ER41306, and that of RNM and BD is supported by the
National Science Foundation grant No. PHY-0652363. We thank S.
Blanchet for useful discussions.

\end{document}